\documentclass{article}

\PassOptionsToPackage{numbers, sort&compress}{natbib}

\usepackage[preprint]{ds_arxiv}

\usepackage[utf8]{inputenc} 
\usepackage[T1]{fontenc}    
\usepackage{hyperref}       
\usepackage{url}            
\usepackage{booktabs}       
\usepackage{amsfonts}       
\usepackage{nicefrac}       
\usepackage{microtype}      
\usepackage{xcolor}         
\usepackage{graphicx}
\usepackage{float}
\usepackage{todonotes}
\usepackage{fancyvrb}

\title{EdgeWisePersona: A Dataset for On-Device User Profiling from Natural Language Interactions}

\author{%
  Patryk Bartkowiak \quad Michal Podstawski\\
  TCL Research Europe \\
  Grzybowska 5A \\
  00-132 Warsaw, Poland \\
  \texttt{\{name.surname\}@tcl.com} \\
  \\
\href{https://huggingface.co/datasets/TCLResearchEurope/EdgeWisePersona}{https://huggingface.co/datasets/TCLResearchEurope/EdgeWisePersona}
\\
\href{https://github.com/TCLResearchEurope/EdgeWisePersona}{https://github.com/TCLResearchEurope/EdgeWisePersona}
}

\begin{document}

\maketitle

\begin{abstract}

This paper introduces a novel dataset and evaluation benchmark designed to assess and improve small language models deployable on edge devices, with a focus on user profiling from multi-session natural language interactions in smart home environments. At the core of the dataset are structured user profiles, each defined by a set of routines - context-triggered, repeatable patterns of behavior that govern how users interact with their home systems. Using these profiles as input, a large language model (LLM) generates corresponding interaction sessions that simulate realistic, diverse, and context-aware dialogues between users and their devices.

The primary task supported by this dataset is profile reconstruction: inferring user routines and preferences solely from interaction history. To assess how well current models can perform this task under realistic conditions, we benchmarked several state-of-the-art compact language models and compared their performance against large foundation models. Our results show that while small models demonstrate some capability in reconstructing profiles, they still fall significantly short of large models in accurately capturing user behavior. This performance gap poses a major challenge - particularly because on-device processing offers critical advantages, such as preserving user privacy, minimizing latency, and enabling personalized experiences without reliance on the cloud. By providing a realistic, structured testbed for developing and evaluating behavioral modeling under these constraints, our dataset represents a key step toward enabling intelligent, privacy-respecting AI systems that learn and adapt directly on user-owned devices.

\end{abstract}

\section{Introduction}

In recent years, the field of artificial intelligence (AI) has witnessed a significant shift toward models that can be developed and executed directly on edge devices, particularly smartphones, TVs, and tablets. This trend has been driven by rapid advancements in both AI model optimization techniques \cite{yu2024edgellmenablingefficientlarge, sander2025acceleratingedgeaioptimizing, saha2025visiontransformersedgecomprehensive} and the computational capabilities of mobile hardware \cite{jayanth2024benchmarkingedgeaiplatforms, li2025largelanguagemodelinference, barker2025advancementsmobileedgecomputing}. Modern mobile devices are now equipped with increasingly powerful processors, dedicated neural processing units (NPUs) \cite{tan2021deeplearningmobiledevices, xu2024fastondevicellminference}, and high-efficiency memory systems, enabling them to run sophisticated AI workloads locally. Simultaneously, the development of compact, efficient models - especially in the realm of generative AI - has accelerated at an unprecedented pace \cite{Wang_2025}. Recent releases such as Qwen \cite{qwen2025qwen25technicalreport}, Phi \cite{abdin2024phi4technicalreport}, Gemma \cite{gemmateam2025gemma3technicalreport}, and other small-scale generative models have demonstrated that meaningful on-device inference is not only possible but increasingly practical for real-world applications. Together, these parallel progressions are catalyzing a fast-paced evolution in mobile AI, bringing once cloud-dependent capabilities into users' hands in real time.

The migration of generative AI capabilities from centralized cloud infrastructures to edge devices brings several compelling advantages \cite{wang2025optimizingedgeaicomprehensive, navardi2025genaiedgecomprehensivesurvey}. Chief among them is reduced latency: by performing inference directly on-device, users experience faster response times without the delays associated with network communication \cite{ye2025jupiterfastresourceefficientcollaborative, tang2025scalingondevicegpuinference, kim2024he2cholisticapproachallocating}. This is particularly beneficial for interactive applications like real-time translation, voice assistants, and on-the-fly content generation. Additionally, enhanced privacy and data security become inherent features of on-device processing, as sensitive user data no longer needs to be transmitted to remote servers \cite{Aminifar_2024}. This local-first approach mitigates risks related to data breaches, unauthorized access, and compliance with data protection regulations. Moreover, on-device AI supports greater reliability and autonomy, allowing applications to function seamlessly in offline or low-connectivity environments. Finally, this paradigm shift also promotes energy efficiency at scale, reducing the load on data centers and cutting down the environmental footprint associated with cloud-based AI processing \cite{mao2024greenedgeaicontemporary, Shafique_2021}. Together, these benefits make a strong case for the continued push toward edge-native generative intelligence.

These emerging on-device capabilities are particularly promising for the development of intelligent, privacy-preserving smart home systems. With generative AI models running locally, it becomes feasible to create highly personalized environments - homes that adapt dynamically to user preferences, routines, and contexts without relying on cloud infrastructure. A crucial component of such systems is the ability to construct and maintain detailed user profiles that inform device behavior over time \cite{wu2024understandingroleuserprofile, suman2022personalizationuserpreferencespartially,balog2025usersimulationeragenerative, birkmose2025ondevicellmshomeassistant}. On-device processing offers a unique advantage here: it enables the continuous refinement of these profiles while keeping sensitive data entirely within the user’s control. However, despite the growing potential, there is a notable absence of publicly available datasets that could accelerate research and development in this space. Existing datasets rarely include rich, structured user profiles or realistic interaction patterns representative of smart home use cases. To address this gap, our work introduces a comprehensive dataset that includes both well-defined, structured user profiles and simulated sessions of user–device interactions grounded in those profiles. Beyond the dataset itself, we also evaluate the capacity of current generative AI models to reconstruct accurate user profiles based solely on observed interaction histories - providing insights into their potential for personalization, as well as their limitations.

\section{Related Work}

As outlined, enabling intelligent, privacy-preserving smart home systems requires constructing structured user profiles through continuous, multi-session interactions. While there has been considerable progress in dialogue systems, preference modeling, and smart home control, existing datasets do not fully meet the requirements for progressive, session-spanning user profile reconstruction. Most available resources focus on isolated dialogues, one-off preference statements, or open-domain conversations without persistent user identity, structured numeric profiling, or domain-specific smart home interactions.

\textbf{PrefEval} \cite{zhao2025llmsrecognizepreferencesevaluating} evaluates models' ability to follow user preferences but consists only of isolated preference-query pairs with no prolonged, multi-session dialogues. Profiles are limited to single preferences rather than structured behavioral representations. In contrast, our dataset supports incremental profile construction across coherent sessions per user.

\textbf{AllenAI SODA} \cite{kim2022soda} provides large-scale synthetic dialogues emphasizing commonsense reasoning in standalone social scenarios. However, it lacks persistent user identities and does not model cumulative preference acquisition or smart home control, which are central to our work.

\textbf{ParlAI PersonaChat} \cite{miller2017parlai} introduces simple personas for chit-chat but restricts each to a single conversation without iterative updates. It focuses on general interests rather than device control or numeric profiling, differing from our emphasis on structured, evolving smart home profiles.

\textbf{LMSYS-Chat-1M} \cite{zheng2024lmsyschat1mlargescalerealworldllm} compiles a vast set of independent user–assistant dialogues across diverse topics. While extensive, it lacks user continuity across sessions and does not focus on domain-specific profile accumulation, unlike our design for persistent, profile-driven smart home interactions.

\textbf{Home-Assistant-Requests} \cite{acon96_home_assistant_requests} addresses smart home commands but only as single-turn queries, without multi-turn dialogues or user profiling. Our dataset extends beyond single commands to simulate realistic, long-term smart home personalization.

\textbf{OpenAssistant/oasst1} \cite{köpf2023openassistantconversationsdemocratizing} offers rich, open-domain conversational data without structured user profiles, session continuity, or domain-specific focus. Unlike our dataset, it does not target behavior modeling across multiple smart home sessions.

In summary, while existing resources contribute to dialogue research, preference modeling, and smart home command understanding, none provide the combination of persistent user identity, multi-session continuity, structured numeric profiling, and smart home domain focus that our dataset offers. By supporting the incremental reconstruction of user profiles over extended, realistic interactions, our work bridges a critical gap between dialogue systems, explainable AI, and behavior-based personalization. 

\section{Dataset Design}

The core component of the dataset consists of natural language sessions between users and their smart home systems. These dialogues simulate realistic, free-form interactions in which users express commands, preferences, or queries. The sessions are grounded in underlying formalized behavioral routines. These routines, along with the user profiles they compose, are also included in the dataset as ground truth annotations. The primary goal is to support and evaluate the ability of compact language models, running directly on edge devices, to reconstruct these routines from unstructured conversational history.

The dataset is synthetic, generated using the DeepSeek-V3 \cite{deepseekai2025deepseekv3technicalreport} language model, but has been meticulously designed according to a set of behavioral and contextual rules described below to ensure realism and consistency. It is inherently self-validating: since sessions are derived from known user routines, and the benchmark task is to recover these routines from the generated dialogues, strong performance by large language models confirms internal coherence and task alignment. In addition, each sample was reviewed by human annotators to ensure also local consistency and fidelity to the intended user behavior. Together, these properties make the dataset a reliable foundation for developing and evaluating lightweight, privacy-preserving language models for on-device personalization.

\subsection{Definitions}

Below we define the core concepts and structures that form the foundation of the dataset. The corresponding data format is provided in the Appendix~\ref{app:schemes}.

The smart home environment in our dataset is characterized by a set of predefined \textbf{device states}, representing the status of household devices at any given time. The modeled devices include a TV (volume, brightness, input source), an air conditioning (AC) unit (temperature, mode, fan speed), a lighting system (brightness, color temperature, operating mode), a speaker (volume, equalizer settings), and a security system (armed status, alarm volume). These baseline states provide the context upon which user interactions operate. Through natural language sessions, users express preferences and commands that result in actions - specific modifications to these device states.

\textbf{Contextual conditions} describe the external and environmental factors that influence user behavior and device interaction patterns. These include temporal, environmental, and situational parameters such as the time of day (morning, afternoon, evening, or night), day of the week, sun phase (before sunrise, daylight, or after sunset), weather conditions (sunny, cloudy, rainy, or snowy), and outdoor temperature. These contextual features serve as inputs to the session generation process and play a central role in determining when routines are activated.

\textbf{Triggers} are specific combinations of contextual conditions that determine when a routine is activated. They represent precise environmental or temporal states that must be met for a routine to apply. For example, a trigger might be configured to activate when the time of day is "evening", the weather is "rainy", and the outdoor temperature is "warm". Triggers thus function as logical match rules that map specific context snapshots to the execution of user behaviors.

\textbf{Routine} refers to a repeatable, structured pattern of user behavior that is activated under specific contextual triggers and results in predefined modifications to smart home device states. Routines encapsulate users' habitual interactions, such as adjusting lighting, setting temperatures, or controlling appliances in response to particular environmental or temporal conditions. Each routine is defined by a pair of components: the \textbf{triggers} that specify when it should activate, and the \textbf{actions} that describe how devices should respond. A collection of routines associated with a user forms a structured behavioral profile, characterizing their typical preferences and interactions within the smart home environment. Determining these routines and reconstructing user profiles solely from natural language interaction sessions is the central goal of the benchmark task defined on top of our dataset.

\subsection{Dataset Construction}

\begin{figure}[H]
\includegraphics[width=\textwidth]{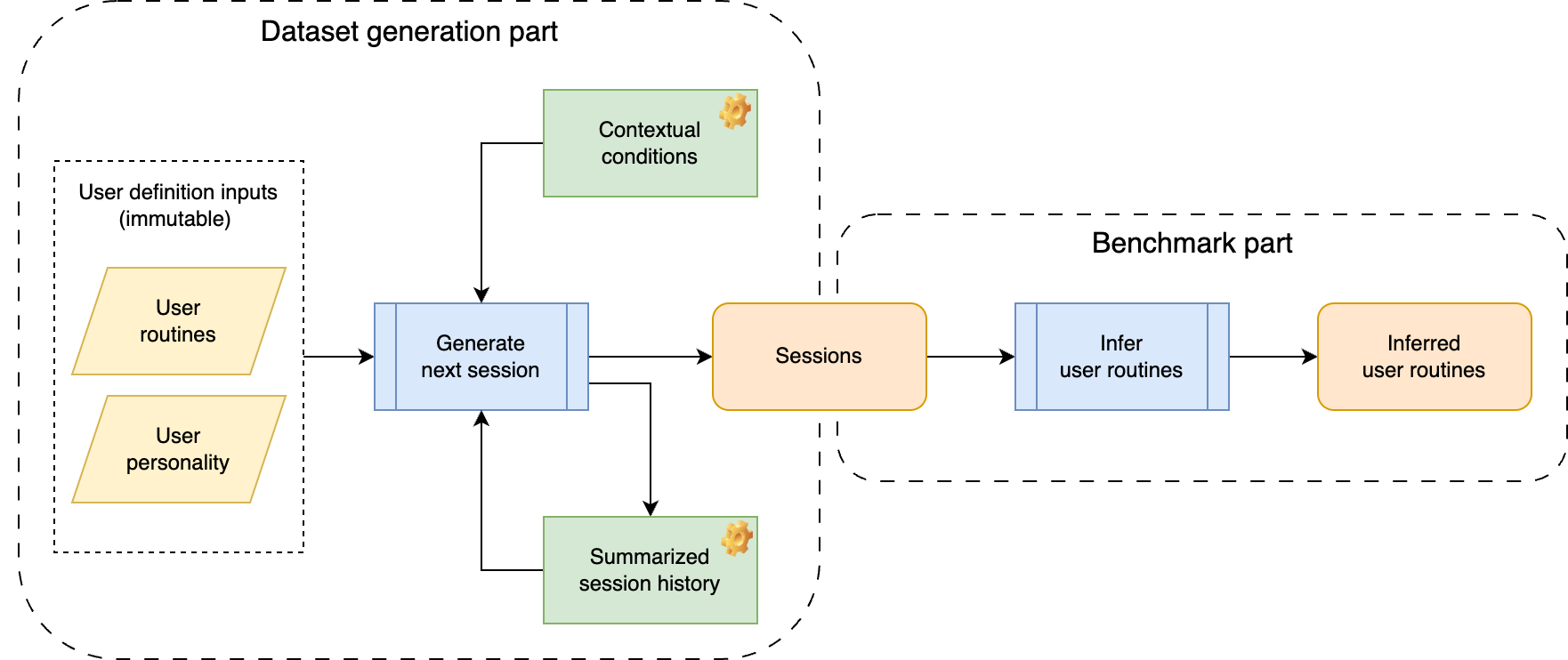}
\caption{\textbf{Architecture of the dataset generation and benchmarking pipeline.} The process is divided into two main parts: dataset generation (left) and benchmarking (right). Yellow boxes represent immutable input definitions: user routines and personality traits. Green boxes indicate dynamic values. Blue boxes denote models processing responsible for generating new sessions or inferring routines. Orange boxes correspond to outputs: generated sessions and inferred user routines. In the benchmark phase, models are evaluated on their ability to reconstruct structured user routines from the generated session data.}
\label{design-figure}
\end{figure}

The dataset is constructed through a controlled, multi-stage process designed to generate realistic, diverse, and structured natural language interactions grounded in user behavior. Each session is derived from a predefined user profile consisting of routines, with the ultimate goal of enabling evaluation of compact language models for routine and profile reconstruction under edge-device constraints. The general architecture of the dataset generation process is illustrated in Figure~\ref{design-figure}.

\textbf{Profile and Routine Specification.} Each user is associated with a profile that combines a set of behavioral routines with a brief textual description of their personality and lifestyle. While this description does not influence the logic of the routines, it is used during session generation to guide the language model toward producing more natural and individualized dialogue. In contrast, the routines determine the behavioral content by specifying which actions are likely to be taken under given contextual conditions. Together, these components ensure that each generated session reflects both consistent behavior patterns and varied, user-specific linguistic style.

\textbf{Contextual Sampling.} For each session, a realistic context is sampled, including variables such as time, day, sun phase and weather conditions. This context determines which routines may be active and influences the content of the generated dialogue.

\textbf{Session Generation.} Given a user profile and sampled context, a natural language interaction is generated. The prompt includes structured profile information, context, and a system message instructing the model to generate a coherent dialogue between the user and their smart home assistant. To promote interaction diversity and avoid repetition, the prompt also includes a summary of previous sessions along with an explicit instruction to generate a new, distinctive dialogue. Sessions vary in routine-dependence; most are grounded in predefined behaviors, while others are spontaneous to reflect natural variation in user interaction.

\textbf{Human Review.} All generated sessions are reviewed by human annotators to ensure local coherence, alignment with the intended user profile, and general fluency. Minor corrections are made when needed, but the generative output is preserved to reflect realistic model behavior.

\textbf{Dataset Structure.} The dataset is organized at the user level, with each user represented across three aligned \texttt{.jsonl} files: one containing their personality description (\texttt{characters.jsonl}), one listing their routines (\texttt{routines.jsonl}), and one storing their interaction sessions (\texttt{sessions.jsonl}). This modular structure supports both supervised training and benchmark evaluation.


The three files are strictly line-aligned: the $n$-th line in each file corresponds to the same user. This design ensures that each user’s personality, routines, and session history are consistently linked. This structure makes the dataset easy to parse and integrate into training or evaluation pipelines while maintaining a clean and modular format.

\subsection{Dataset Statistics}

The dataset includes interactions from 200 users, with 50 natural language sessions collected for each user (10000 sessions in total). Each session reflects dialogue with the smart home system, supporting the extraction and inference of structured routines. Each session is associated with at most one routine, with approximately 75\% of sessions reflecting a routine-driven behavior and the remaining 25\% representing spontaneous or routine-free interactions. This distribution is intended to mirror realistic user behavior, where not every interaction follows a habitual pattern. A summary of key dataset statistics is provided in Table~\ref{tab:session-stats}.

\begin{table}[H]
\centering
\caption{Overview of key statistics for the dataset.}
\label{tab:session-stats}
\begin{tabular}{p{0.5\textwidth} p{0.1\textwidth}}
\toprule
\textbf{Key} & \textbf{Value} \\
\midrule
\# Users & 200 \\
\# Sessions per User & 50 \\
\# Total sessions & 10000 \\
\midrule
Avg. \# Messages per Session & 9.88 \\
Avg. \# Routines per User & 3.98 \\
Avg. \# Triggers per Routine & 4.98 \\
Avg. \# Devices per Routine & 3.11 \\
Avg. \# Actions per Routine & 8.24 \\
\bottomrule
\end{tabular}
\end{table}

\section{Benchmark Results}

Our evaluation compares state-of-the-art large LLMs with smaller, resource-efficient models, focusing on their long-context capabilities and overall accuracy. Model selection was guided by performance and context length constraints, using comparative metrics from the LLM-Stats leaderboard \cite{llmstats_github_2024}. As edge-deployable candidates, we selected top-performing models with fewer than 5B parameters that are compatible with mobile and resource-constrained environments, and that support extended input contexts of at least 128k tokens - necessary for handling the long, multi-turn sessions in our dataset. The models used in our evaluation include \textbf{Gemma-3-4B} \cite{gemmateam2025gemma3technicalreport}, \textbf{Qwen-2.5-3B} \cite{qwen2025qwen25technicalreport}, \textbf{Llama3.2-3B} \cite{grattafiori2024llama3herdmodels, meta2024llama32}, and \textbf{Phi-4-mini-4B} \cite{abdin2024phi4technicalreport}, all of which offer strong accuracy relative to their size.

As the foundation for our baseline comparisons, we include state-of-the-art large language models, such as \textbf{GPT-4o} \cite{openai2024gpt4technicalreport, openai2024gpt4ocard}, \textbf{Gemini-2.5-Flash} \cite{geminiteam2025geminifamilyhighlycapable, google2025gemini25flash}, and \textbf{DeepSeek-V3} \cite{deepseekai2025deepseekv3technicalreport}. These models represent the upper bound in terms of both scale and capability, and serve as strong reference points for evaluating the performance of smaller, mobile-deployable models.

\paragraph{Compute setup.} Large-scale models were accessed via their official hosted APIs. Compact models were benchmarked locally on a single NVIDIA RTX 3090 GPU (24 GB). To validate on-device compatibility, we exported each model to mobile-friendly formats and tested execution across four runtime stacks, confirming that they can be deployed and run on real edge hardware. We used three smartphones for this evaluation: the Qualcomm Snapdragon 675-based TCL 10 5G and TCL T1 Pro, and the MediaTek Dimensity 9300-powered OPPO Find X7. Execution was tested using widely adopted mobile inference frameworks: Alibaba MNN \cite{alibaba2020mnn}, Executorch \cite{executorch} with MediaTek NeuroPilot \cite{neuropilot}, ONNX Runtime \cite{onnxruntime} Mobile with NNAPI and GPU acceleration, and llama.cpp \cite{llamacpp} for CPU-only fallback.

\subsection{Evaluation Protocol}

We evaluate each model under a uniform prediction-count policy to ensure fair comparison. Specifically, models are prompted to predict exactly the same number of routines as the ground truth contains. If a model generates more routines than the reference, we truncate the extra predictions; if it generates fewer, we retain the given predictions (leaving the missing routines as blank). This alignment guarantees a one-to-one correspondence for scoring each predicted routine against its true counterpart. All accuracy metrics are then computed per routine and averaged over the corpus.


\subsection{Routine-Level Accuracy}

We report two complementary metrics for overall routine prediction: strict exact-match accuracy and Jaccard similarity. \textbf{Exact-match accuracy} measures the proportion of predicted routines that match the ground-truth routines exactly - across all fields, including contextual triggers, actions, and their parameters. This metric provides a stringent assessment of model performance, as even a small deviation results in a non-match. \textbf{Jaccard similarity}, by contrast, measures the degree of partial overlap between the predicted and reference sets of routine components, offering a more forgiving view of routine reconstruction quality. We express both metrics as percentages to support clear interpretation and facilitate direct comparison between models.

\begin{figure}[H]
\includegraphics[width=\textwidth]{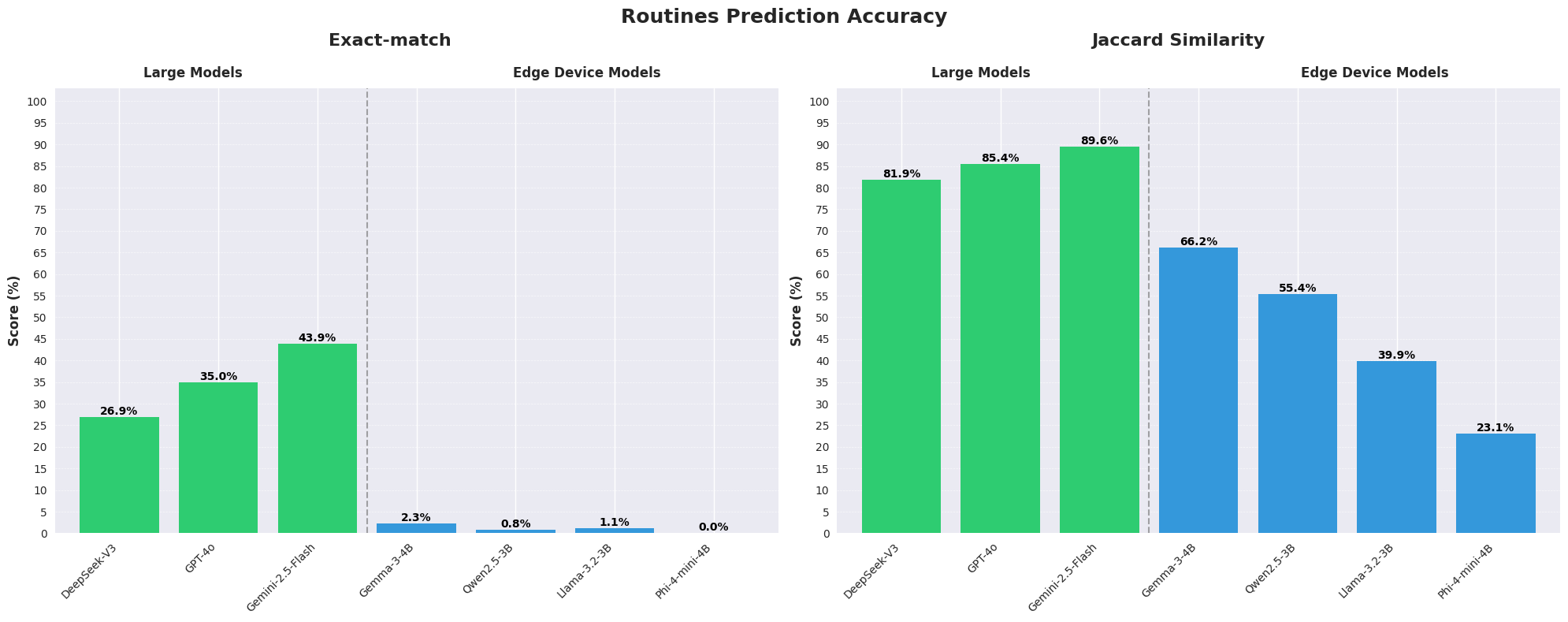}
\caption{Comparison of model performance on the routine prediction task using exact-match accuracy (left) and Jaccard similarity (right). Results are grouped by model size. Exact-match reflects full structural correctness, while Jaccard similarity captures partial overlap between predicted and reference routines.}
\label{benchmark-fig1}
\end{figure}

As shown in Figure~\ref{benchmark-fig1}, large models substantially outperform smaller ones on both metrics. For example, Gemini-2.5-Flash achieves an exact-match accuracy of 44\%, followed by GPT-4o at 35\% and DeepSeek-V3 at 27\%. In comparison, the best-performing small model, Gemma-3-4B, achieves only 2\%, while the remaining small models are close to zero - indicating they rarely predict a complete routine correctly.

A similar trend appears in Jaccard similarity scores: Gemini reaches 90\%, followed by GPT-4o (85\%) and DeepSeek-V3 (82\%). Among the smaller models, Gemma-3-4B performs best with a score of 66\%, while Qwen-2.5-3B and Llama3.2-3B achieve 55\% and 40\% respectively. Phi-4-mini-4B lags behind with 23\%. These results highlight a clear performance gap: while large models often capture most aspects of a routine (reflected in high Jaccard similarity scores), smaller models struggle to do so - and achieving perfect predictions remains a challenge even for the best systems.


\subsection{Trigger vs. Action Prediction}

\begin{figure}[H]
\includegraphics[width=\textwidth]{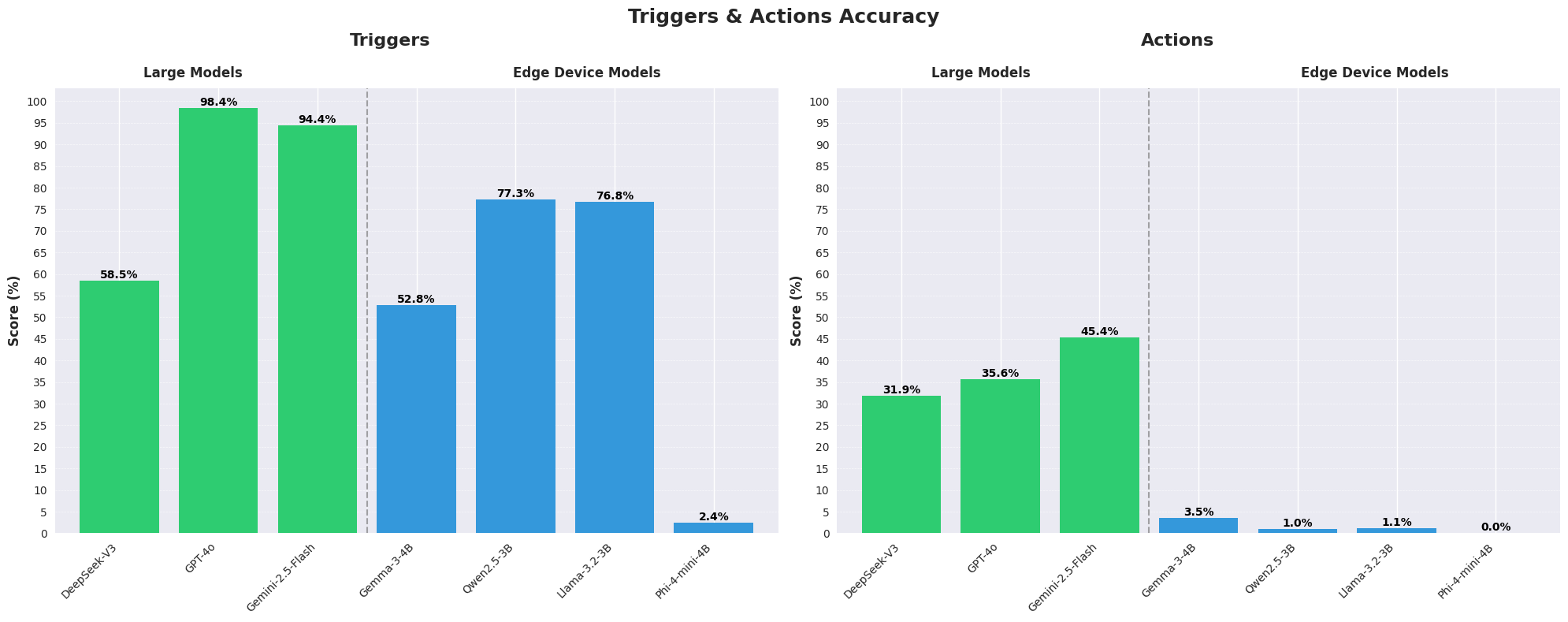}
\caption{Model accuracy on trigger (left) and action (right) prediction. Scores are computed using exact-match evaluation against the best-matching ground-truth routine.}
\label{benchmark-fig2}
\end{figure}

We further break down model performance by separately evaluating predictions of triggers and actions. For each routine predicted by a model, we first identify the most similar ground-truth routine using Jaccard similarity. We then apply exact-match evaluation to the trigger and action components of that matched pair - checking whether each part was reproduced exactly. As shown in Figure~\ref{benchmark-fig2}, this reveals a stark difference in difficulty between the two.

All models perform substantially better on trigger prediction than on action prediction. The large models almost always identify the correct trigger: for example, GPT-4o and Gemini-2.5-Flash each exceed 90\% trigger accuracy, and DeepSeek-V3 reaches 58\%. Even smaller models like Qwen-2.5-3B and Llama3.2-3B achieve 77\%, suggesting that compact models can often guess the correct initiating condition. This is likely because triggers consist of categorical values (e.g., time of day, sun phase), which are easier to recover than scalar outputs.

In contrast, action prediction proves far more challenging. The best-performing large model (Gemini-2.5-Flash) achieves only about 45\% accuracy on full action sequences, followed by GPT-4o at 36\% and DeepSeek-V3 at 32\%. These results indicate that even advanced models struggle to reproduce full sets of actions precisely. For the smaller models, performance is near-zero: Gemma-3-4B reaches just 4\%, while the others fail to generate correct actions in most cases. This highlights the structured complexity of action prediction as a major bottleneck in accurate routine reconstruction.

\subsection{Scalar Fields Accuracy}

Many routine actions involve scalar parameters (numeric values such as brightness levels, volume settings, or target temperatures). We evaluate each model’s ability to predict these numeric fields using mean absolute error (MAE). \autoref{tab:mae-exact} reports the results. The large models deliver near-perfect precision: GPT-4o achieves an average MAE of approximately 0.014, and Gemini-2.5-Flash is nearly identical at 0.018. DeepSeek-V3 is slightly higher but still low, at 0.13. In contrast, compact models miss the mark by an order of magnitude. Gemma-3-4B reaches 1.049, while Qwen-2.5-3B and Llama3.2-3B follow with 1.837 and 1.895, respectively. Phi-4-mini-4B performs worst, with an average MAE of 7.348 (peaking at 13.399 on the lights brightness field). These gaps show that large LLMs can infer numeric device settings with sub-unit accuracy, whereas edge-oriented models frequently deviate by several units - well outside a tolerable range for reliable smart-home control.

\begin{table}[H]\centering
\caption{Mean Absolute Error (MAE) for predicted scalar device fields. These fields include numeric settings for various devices in the routines. Lower values denote more accurate predictions of user-specific settings. Values are averaged across all users.}
\footnotesize
\label{tab:mae-exact}
\resizebox{\textwidth}{!}{%
\begin{tabular}{lccccccc}
\toprule
\textbf{Model} & \textbf{Avg. MAE} & \textbf{TV Volume} & \textbf{TV Brightness} & \textbf{AC Temp.} & \textbf{Lights Bright.} & \textbf{Speaker Vol.} & \textbf{Security Alarm} \\
\midrule
DeepSeek-V3      & 0.130 & 0.087 & 0.138 & 0.030 & 0.340 & 0.064 & 0.119 \\
GPT-4o           & \textbf{0.014} & \textbf{0.017} & \textbf{0.017} & \textbf{0.000} & 0.052 & \textbf{0.000} & \textbf{0.000} \\
Gemini-2.5-Flash & 0.018 & \textbf{0.017} & 0.034 & 0.008 & \textbf{0.026} & 0.025 & \textbf{0.000} \\
\midrule
Gemma-3-4B       & \textbf{1.049} & \textbf{1.172} & 1.970 & \textbf{0.158} & \textbf{1.680} & \textbf{0.266} & \textbf{1.050} \\
Qwen2.5-3B       & 1.837 & 1.436 & 1.702 & 0.422 & 4.867 & 1.355 & 1.242 \\
LLaMA-3.2-3B     & 1.895 & 2.245 & \textbf{1.556} & 0.258 & 3.773 & 0.702 & 2.838 \\
Phi-4-mini-4B    & 7.348 & 6.250 & 5.833 & 1.327 & 13.399 & 5.661 & 11.616 \\
\bottomrule
\end{tabular}%
}
\end{table}

\subsection{Categorical Fields Accuracy}
Finally, we assess \textbf{categorical fields} in the routines, such as device modes or preset names (e.g.\ \textit{cool}, \textit{heat} or \textit{auto} as AC mode settings). For these discrete settings, we measure exact-match accuracy - whether the model selected the correct category out of all possible options.  \autoref{tab:class-accuracy-exact} summarizes results across seven representative fields. The large models get almost every categorical choice right. GPT-4o is the top performer, averaging close to 97\% accuracy across all fields, followed by DeepSeek-V3 and Gemini-2.5-Flash at approximately 95\% and 92\%, respectively. Among compact models, Gemma-3-4B does best, attaining almost 88\% accuracy - respectable, yet still below the large models. The other edge-sized models struggle: Qwen2.5-3B is correct a little over 50\% of the time, while Phi-4-mini-4B and LLaMA-3.2-3B trail at roughly 32\% and 24\%, respectively. These disparities underscore how much fine-grained device settings rely on nuanced contextual cues: large LLMs capture those cues consistently, whereas smaller models often default to common or \textit{safe} categories.  For reliable prediction of categorical device parameters in real deployments, the larger models therefore retain a clear advantage.

\begin{table}[H]\centering
\caption{Accuracy for categorical device fields in the routines. Each column corresponds to a device setting that the model must predict correctly. The values indicate the percentage of predictions that exactly matched the true category.}
\footnotesize
\label{tab:class-accuracy-exact}
\resizebox{\textwidth}{!}{%
\begin{tabular}{lcccccccc}
\toprule
\textbf{Model} & \textbf{Avg. Acc.} & \textbf{TV Input} & \textbf{AC Mode} & \textbf{AC Fan} & \textbf{Lights Color} & \textbf{Lights Mode} & \textbf{Speaker EQ} & \textbf{Sec. Armed} \\
\midrule
DeepSeek-V3      & 95.2 & 96.0 & 98.9 & 98.3 & 99.2  & \textbf{94.6} & 91.3 & 88.1 \\
GPT-4o           & \textbf{96.9} & \textbf{98.3} & \textbf{99.4} & \textbf{99.2} & \textbf{100.0} & 90.1 & \textbf{97.9} & \textbf{93.6} \\
Gemini-2.5-Flash & 91.8 & \textbf{98.3} & 99.2 & 98.6 & 99.2  & 66.1 & 93.7 & 87.8 \\
\midrule
Gemma-3-4B       & \textbf{87.7} & \textbf{84.7} & \textbf{94.7} & \textbf{94.1} & \textbf{94.6}  & \textbf{65.2} & \textbf{93.7} & \textbf{87.2} \\
Qwen2.5-3B       & 56.0 & 27.9 & 76.2 & 75.1 & 78.8  & 58.6 & 50.1 & 25.6 \\
LLaMA-3.2-3B     & 23.2 & 14.7 & 26.7 & 25.8 & 25.7  & 24.8 & 26.4 & 19.6 \\
Phi-4-mini-4B    & 32.2 & 23.1 & 27.2 & 41.9 & 60.9  & 13.2 & 41.0 & 17.8 \\
\bottomrule
\end{tabular}%
}
\end{table}

\subsection{Open Source Benchmarking Tools}


The full benchmarking pipeline, including evaluation scripts and data class definitions, is open-source and publicly available. This allows other researchers to reproduce our results, extend the benchmark to new models, or adapt it for related user profiling tasks. The extensibility of the framework supports further exploration into long-context modeling, explainable personalization, and low-latency edge deployment of conversational AI.

\section{Limitations}

While our dataset provides a valuable testbed for evaluating the ability of language models to reconstruct structured user profiles in smart home settings, several limitations of the dataset design and scope should be noted.

First, the evaluation of model performance is focused primarily on reconstructing structured profiles from predefined interaction sessions. This approach does not yet address challenges related to long-term personalization, profile drift over time, or handling ambiguous or conflicting user signals.

Second, the current study assumes a high degree of user cooperation and clean data input, which may not always be feasible in practical applications. In real settings, data quality can be affected by noise, interruptions, or inconsistent user engagement, potentially impacting model reliability and the quality of generated profiles.

These limitations highlight key areas for future research aimed at refining system robustness, scalability, and real-world applicability.

\section{Future Work}

The dataset and evaluations presented in this study establish a foundation for developing privacy-preserving, on-device intelligent agents for smart home environments. However, several avenues remain open for further exploration.

One promising direction is the expansion of user profile complexity to incorporate evolving behaviors, multi-user dynamics within shared spaces, and more nuanced contextual factors such as emotional states or environmental changes. This would allow models to learn and adapt to richer, more lifelike scenarios.

Future iterations of this research would also benefit from real-world deployment and feedback loops, where user interaction data - collected and processed entirely on-device - can inform iterative improvements.





\section{Conclusions}

This work presents a novel dataset designed to enable the reconstruction of structured user profiles from natural language interaction sessions with smart home devices. By starting from rich, well-defined profiles and generating realistic sessions using a large language model, the dataset provides a diverse environment for developing and benchmarking compact, on-device AI systems.

While the results demonstrate the feasibility of profile reconstruction using small language models, it is important to acknowledge that current state-of-the-art on-device models still lag behind fully-fledged large language models in their ability to accurately infer complex routines and user behaviors. Significant work remains to close this performance gap and optimize inference capabilities under the computational constraints of edge devices. We believe this resource lays an important foundation for future advances in privacy-preserving personalization, behavior-based user modeling, and the development of transparent, adaptive smart environments directly on user-owned devices.







{\small
\bibliographystyle{ACM-Reference-Format}
\bibliography{ds_arxiv}
}


\appendix


\section{Technical Appendices and Supplementary Material}
\label{app:schemes}

To support transparency, usability, and reproducibility, we include in this appendix detailed schemas representing the internal structure of two key components of our dataset: \texttt{routines} and \texttt{sessions}. These schemas serve as precise technical documentation, defining the data format used for each synthetic user’s behavior and interaction history.

The \texttt{routines} schema provides a full description of how each user’s behavioral logic is encoded. Routines are central to both the data generation and evaluation processes, and their schema captures the nested structure of contextual triggers, associated device actions, and routine-level metadata such as priority. Each routine is composed of clearly defined fields for environmental and temporal conditions (e.g., time of day, weather, sun phase), as well as structured action blocks targeting specific devices (e.g., lighting, AC, speaker). This format supports interpretable comparisons between ground-truth behavior and reconstructed routines during benchmarking.
\\

\begin{Verbatim}[frame=single, fontsize=\small, label=Routine Model Structure]
Routine:
  triggers:
    time_of_day: Optional["morning" | "afternoon" | "evening" | "night"]
    day_of_week: Optional["weekday" | "weekend"]
    sun_phase: Optional["before_sunrise" | "daylight" | "after_sunset"]
    weather: Optional["sunny" | "cloudy" | "rainy" | "snowy"]
    outdoor_temp: Optional["very cold" | "cold" | "mild" | "warm" | "hot"]

  actions:
    tv:
      volume: Optional[int]
      brightness: Optional[int]
      input_source: Optional["HDMI1" | "HDMI2" | "AV" | "Netflix" | "YouTube"]

    ac:
      temperature: Optional[int]  # Range: 16–30
      mode: Optional["cool" | "heat" | "auto"]
      fan_speed: Optional[int]  # Range: 0–3

    lights:
      brightness: Optional[int]
      color: Optional["warm" | "cool" | "neutral"]
      mode: Optional["static" | "dynamic"]

    speaker:
      volume: Optional[int]
      equalizer: Optional["bass boost" | "balanced" | "treble boost"]

    security:
      armed: Optional[bool]
      alarm_volume: Optional[int]
\end{Verbatim}

The \texttt{sessions} schema defines the structure of the generated natural language dialogues. Each session entry encapsulates a complete multi-turn interaction between a user and their smart home system, enriched with metadata about the contextual state in which the session takes place. The schema captures the contextual snapshot and optional markers for routine activation or model-generated reasoning. This structured format allows for controlled evaluation scenarios and enables models to learn associations between language, context, and user behavior.
\\

\begin{Verbatim}[frame=single, fontsize=\small, label=Session Model Structure]
Session:
  session_id: int

  meta:
    time_of_day: "morning" | "afternoon" | "evening" | "night"
    day_of_week: "weekday" | "weekend"
    sun_phase: "before_sunrise" | "daylight" | "after_sunset"
    weather: "sunny" | "cloudy" | "rainy" | "snowy"
    outdoor_temp: "very cold" | "cold" | "mild" | "warm" | "hot"

  messages:
    - role: "user" | "assistant"
      text: str

  applied_routines: List[int]
\end{Verbatim}

Together, these schemas provide a clear and comprehensive preview of the dataset’s internal design. They are intended to guide parsing and integration into training pipelines and we hope these specifications will facilitate robust adoption and adaptation of the dataset across a wide range of downstream tasks and evaluation settings.

\end{document}